\documentclass{article}
\usepackage{epsf}
\usepackage{amstex}
\usepackage{amssymb}

\begin{document}
\thispagestyle{empty}
\parskip=12pt
\raggedbottom
 
\def\mytoday#1{{ } \ifcase\month \or
 January\or February\or March\or April\or May\or June\or
 July\or August\or September\or October\or November\or December\fi
 \space \number\year}
\noindent
\hspace*{9cm} BUTP/2000-15\\
\vspace*{1cm}
\begin{center}
{\LARGE Unexpected results in asymptotically free quantum field theories }
\footnote{Work supported in part by Schweizerischer Nationalfonds.}
 
\vspace{1cm}
Peter Hasenfratz and Ferenc Niedermayer\footnote{On leave from 
the Institute of Theoretical Physics, E\"otv\"os University, Budapest}
\\
Institute for Theoretical Physics \\
University of Bern \\
Sidlerstrasse 5, CH-3012 Bern, Switzerland

\vspace{0.5cm}

\nopagebreak[4]
 
\begin{abstract}
We study the behavior of asymptotically free (AF) spin and gauge
models when their continuous symmetry group is replaced by different
discrete non-Abelian subgroups. Precise numerical results with relative
errors down to O(0.1\%) suggest that the models with large subgroups
are in the universality class of the underlying original models. 
We argue that such a scenario is consistent with the known properties 
of AF theories. The small statistical errors allow a detailed 
investigation of the cut-off effects also. At least up to correlation 
lengths $\xi\approx 300$ they follow effectively an $\propto a$ rather 
than the expected $\propto a^2$ form both in the O(3) and in 
the dodecahedron model.
\end{abstract}
 
\end{center}
\eject

\section{Introduction}

AF quantum field theories are relevant both in particle and in condensed
matter physics. Their standard formulation is based on a global ($d=2$ spin
models), or local ($d=4$ gauge theories) non-Abelian continuous symmetry
group. In this work we discuss a scenario which, beyond its field theoretical
interest, might open new analytic and numerical ways to study such theories.

Lattice regularization allows (both in spin and gauge models) 
to replace the continuous 
symmetry group by one of its finite discrete non-Abelian subgroups. The new
model has a reduced symmetry and, due to the discreteness of the group, a
finite action gap. If the bare coupling constant $g$ is large, then the
fluctuations are large and these new features of the model will have a small
effect. 
At small $g$, however, the finite action gap kills the fluctuations and
the system will be frozen. We expect therefore a phase transition
between the strong and weak coupling regions - possibly even two transitions,
if there is a massless phase between them. As opposed to that, the original AF
model is expected to have a single phase only approaching the continuum limit
as $g \rightarrow 0$.

Denote the phase transition point at the end of the strong coupling phase 
of the discrete subgroup model by
$g_c$ and assume that the transition is second order. (We shall discuss the
case of a first order transition later.) 
We suggest that the quantum field theory obtained in the limit $g \searrow g_c$
is in the same universality class as the original AF model 
if the subgroup is sufficiently large.

In the early 80's gauge models on discrete non-Abelian subgroups were
introduced as an {\it approximation} to the SU(2) Yang-Mills theory
\cite{Rebbi}. 
That time it was considered as obvious that this 
is an approximation only since the
Wilson loop expectation values measured at some coupling $g$ were
different. Our suggestion above, however, refers to the situation when 
the discrete and continuous group models are compared at the same
large correlation length (approaching the continuum limit) 
for physical quantities.

A nice example, where a discrete internal symmetry of the action is
elevated to a continuous symmetry, is the d=2 XY model. Considering 
perturbations which break the O(2) symmetry down to Z($N$) in the
massless low temperature phase, for $g>g(N)$
the discrete symmetry is elevated to
full O(2) in the long distance predictions if $N>4$ \cite{Jose}. Our
case is more complicated since AF theories have a non-perturbative
finite mass gap.

We were inspired to investigate models with discrete subgroups 
through the observations made by Patrascioiu and Seiler
a few years ago \cite{PS} who noticed that,
within the statistical errors, the physical results of the dodecahedron 
spin model for $g>g_c$ are consistent with that of the O(3) non-linear 
$\sigma$-model. They concluded that the O(3) model should also have 
a phase transition and so, none of the two models is AF. 
We believe that overwhelming evidence exists that the O(3) $\sigma$-model 
is AF and our suggestion implies that the dodecahedron model is also AF.
This statement should be understood in the sense that the
{\em physical} running coupling goes to zero as the corresponding
momentum scale goes to infinity.

If our suggestion is correct, the relation between the bare coupling
of the original model and that of its discrete subgroup version is
non-analytic. The standard arguments which demonstrate the
universality of the first two terms in the
beta function describing the change of the
bare coupling under the change of the cut-off are not valid.

There exists a special class of non-Abelian subgroups, the dihedral 
subgroups of SU(N) which are not considered in this work.
(The discrete subgroups of SU(3) are discussed in Ref.~\cite{DSU3}.)
These subgroups can be made arbitrarily large and, although they do not 
go over to SU(N) in this limit, the action gap gets arbitrarily small. 
Due to the fact that they inherit some of those properties of SU(N) 
which are thought to be relevant for confinement in the Yang-Mills theory, 
Kaplan raised the possibility that they are in the universality class 
of SU(N)~\cite{Kaplan}. It would be interesting to investigate this
scenario also.

\section{AF spin models in $d=2$}

We studied discrete spin models where the symmetry group is one of the
non-Abelian subgroups of O(3). These subgroups are the symmetry
groups of regular polyhedra. Embedding the regular polyhedron in $S_2$ 
the corners define the discrete directions of the unit spin 
${\bf S}(x)$.
We shall refer to the discrete spin model by the name of 
the corresponding polyhedron. For all the subgroups we used 
the standard nearest-neighbor action.

We found that the spin models with small subgroups (tetrahedron, 
octahedron and cube with 4, 6 and 8 directions, respectively) 
are definitely {\it not} equivalent to the O(3) non-linear
$\sigma$-model. 
The two largest subgroups (icosahedron and dodecahedron with 
12 and 20 directions, respectively) 
could not be distinguished from the O(3) $\sigma$-model, however.
In particular, for the dodecahedron the agreement is established on 
the level of O(0.1\%).

It is easy to show that the tetrahedron model is identical to
the $q=4$ Potts model with $\beta_{{\rm tetra}}=3/4
\beta_{{\rm Potts}}^{q=4}$, where $\beta=1/g$.
Similarly, the cube model is the product of 3 independent Ising
models. This is obvious if one takes the 8 vectors pointing 
to the corners of the cube as $1/\sqrt{3}(\pm 1,\pm 1,\pm 1)$. 
Then the cube action becomes a sum of 3 independent Ising actions, 
with the relation $\beta_{\rm cube}=3 \beta_{\rm Ising}$ between 
the corresponding couplings.
A lot is known about the $q=4$ Potts and the Ising model and 
we can certainly exclude that they are in the universality 
class of the O(3) $\sigma$-model. 

We have chosen physical quantities which can be measured precisely:
the finite-size scaling function, the renormalized zero momentum 4-point
coupling $g_{\rm R}$ and the same coupling in a finite physical
volume. 
It is the latter quantity which can be measured with the highest
precision. 

\subsection{The finite-size scaling function}

An $L \times L$ periodic box is considered and the second
moment correlation length $\xi(L)$ is measured. 
The relative change of the correlation length when $L$ is doubled, 
$\xi(2L)/\xi(L)$ is studied as the function of $\xi(L)/L$. 
Both these ratios are physical quantities if $L$ and 
$\xi(L)$ are much larger than the lattice unit $a$. 
The finite-size scaling function is a valuable
tool to connect the small and large volume regimes\cite{FSS}. This
technique has been used very effectively by L\"uscher, Weisz and
Wolff\cite{LWW} in their study of the running coupling in the O(3) model. 
The finite-size scaling function has been measured precisely by Caracciolo et
al.\cite{Car} in the same  model for a large set of $\xi(L)/L$ values, 
using the second-moment correlation length. The latter is defined 
in an $L \times L$ box as
\begin{equation}
\label{xiL}
\xi(L)=\frac{1}{2\sin(\pi/L)}\sqrt{\frac{G_2(0)}{G_2(k_0)}-1}\,,
\end{equation}
where $G_2(k)$ is the 2-point spin correlation function in Fourier
space and $k_0=(2\pi/L,0)$.
The finite-size scaling function was also studied for the $q=4$ 
Potts model (equivalent to the tetrahedron model) by Salas and 
Sokal\cite{Sal}. 
The two functions are completely, even qualitatively
different. The largest deviation between the O(3) and tetrahedron
finite-size scaling curves is around $\xi(L)/L=0.4$. We have decided,
therefore to study this function in the neighborhood of this value
for the different subgroups as the continuum limit is approached. 

\begin{figure}[htb]
\begin{center}
\vskip -3mm
\leavevmode
\epsfxsize=100mm
\epsfbox{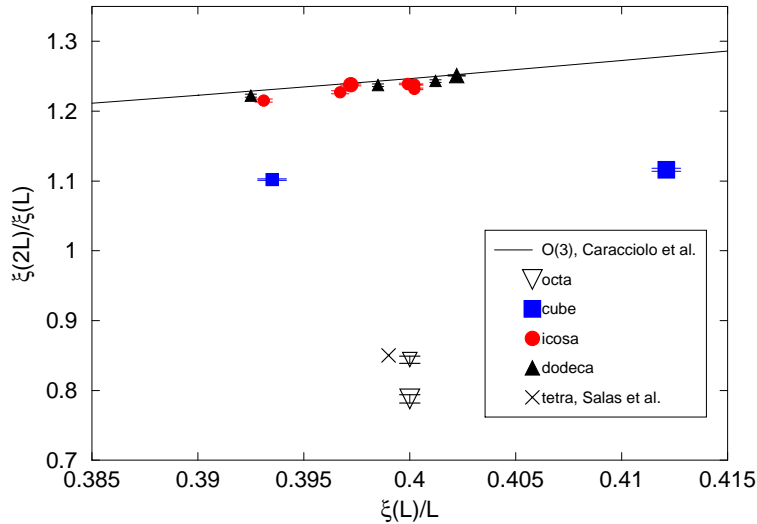}
\vskip -6mm
\end{center}
\caption{{}The $\xi(L)/L \sim 0.4$ part of the finite-size scaling curve (solid
  line) of the O(3) non-linear $\sigma$-model~\cite{Car} is compared with
  the results of discrete spin models based on different non-Abelian subgroups
  of O(3). Points with identical symbols belong to the same subgroup at
  different $\xi(L)$ values. 
  (The largest $\xi$ corresponds to a larger symbol.)
  The points of the icosahedron and dodecahedron
  models move towards the O(3) curve as the correlation length is increased
  and no difference beyond the statistical errors can be seen at the largest
  correlation length.}
\label{fig:fss3}
\end{figure}

In Figure~\ref{fig:fss3} the small neighborhood of $\xi(L)/L=0.4$ 
is shown only. 
The solid line is the O(3) finite-size scaling curve in the
continuum limit \cite{Car}. Points with identical symbols belong 
to the same subgroup at different $L$ values. 
At fixed $\xi(L)/L$ the $L \rightarrow \infty$ limit
is the continuum  limit. The  tetrahedron, octahedron and cube models
deviate strongly from O(3) and the deviation grows as we move
towards the continuum. The icosahedron and dodecahedron approach
the O(3) and at the largest $L$ (increased size symbols) they lie on
the O(3) curve within the statistical errors.

In Figure~\ref{fig:fss3} the points scatter around $\xi(L)/L=0.4$. 
With careful tuning one might stay closer to this value,
but since the scaling function describes physics at every $\xi(L)/L$ 
this was not necessary for the conclusion.

\subsection{The renormalized zero momentum 4-point coupling $g_{\rm R}$}

Define the quantity $g_{\rm R}$ as
\begin{equation}
\label{gR}
g_{\rm R}=\left( \frac{L}{\xi(L)}\right)^2
\left(1+\frac{2}{N}-\frac{\langle({\bf M}^2)^2\rangle}
{\langle{\bf M}^2\rangle^2} \right), 
\end{equation}
where $\xi(L)$ is the second moment correlation length, eq.~(\ref{xiL}),
$z=L/\xi(L)$, $N=3$ for O(3) and ${\bf M}$ is the magnetization,
\begin{equation}
M^a=\sum_x  S^a(x) \,.
\end{equation}
The coupling $g_{\rm R}^\infty$ (which is denoted in the literature
simply by $g_{\rm R}$) is defined as the infinite volume 
($z\to\infty$) limit of $g_{\rm R}$.
(It is assumed, of course, that first the continuum limit
$\xi\to\infty$ is taken for fixed $z$.)
For the O(3) $\sigma$-model $g_{\rm R}^\infty$ has been calculated
using the form-factor bootstrap method: $g_{\rm R}^\infty=6.770(17)$ in
agreement with the MC result $g_{\rm R}^\infty=6.77(2)$~\cite{Bal}.

\begin{figure}[htb]
\begin{center}
\vskip -3mm
\leavevmode
\epsfxsize=100mm
\epsfbox{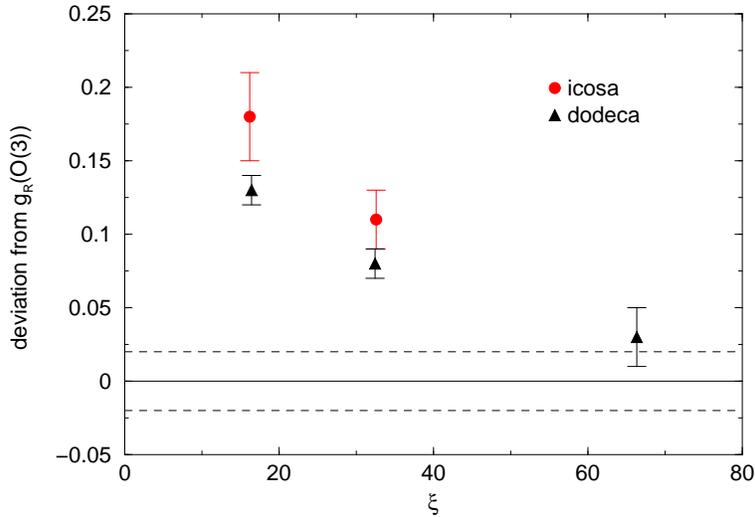}
\vskip -6mm
\end{center}
\caption{{}The deviation between the renormalized coupling of O(3)
and that of the two largest subgroups as the function of the
correlation length.}
\label{fig:gr}
\end{figure}

We have measured $g_{\rm R}$ for the icosahedron and dodecahedron
models for $z\sim 6$ and extrapolated to infinite volume
using the finite size formula of ref.~\cite{Bal}. 
The latter was obtained from the leading $1/N$ expansion\cite{largeN} 
applied to finite volume.
Although it was derived at large $N$, as an empirical formula it
works well also  at N=3~\cite{Bal}.
Figure~\ref{fig:gr} gives the deviation from the O(3) result as 
the function of $\xi$.  ($\xi \rightarrow \infty$ is the continuum limit). 
For comparison, the coupling from
the cubic group is $g_{\rm R}^{{\rm cubic}}=1/3
g_{\rm R}^{{\rm Ising}}$ with $g_{\rm R}^{{\rm Ising}}=14.6975(1)$ from
ref.\cite{Ising}, i.e. for this case the deviation from O(3) is 1.87(2).

As Figures~\ref{fig:fss3} and \ref{fig:gr} show, the two largest 
non-Abelian 
subgroups behave qualitatively differently from the small subgroups. 
The results suggest that the icosahedron and dodecahedron models are 
in the same universality class as the O(3) non-linear $\sigma$-model.

\subsection{A high precision test}

We discuss now a test using a physical quantity which can be measured
both in the O(3) and in the dodecahedron model to very high
precision. 
The coupling $g_{\rm R}$ in eq.~(\ref{gR}) is a physical quantity 
also in a finite physical volume, i.e. in the continuum limit 
keeping $z=L/\xi$ fixed. The numerical problem is much easier at some 
moderate value of $z$ than at large $z$.
We have chosen to measure in the vicinity of the arbitrary value 
$z_0=2.32$.

For O(3) we have used the standard nearest
neighbor (nn) and the parametrized fixed-point (FP) 
actions \cite{HN}, while
for the dodecahedron model the nn action was taken. 

Table~\ref{grdata} summarizes the results for the three actions
considered. The last column, $g_{\rm R}(z_0)$ is derived 
from the previous column as explained below. 

\begin{table}[htbp]
  \begin{center}
\begin{tabular}{|rlllll|}
\hline
$L$ & \quad $\beta$ &  \quad $z$  & \quad $g_{\rm R}$ & $g_{\rm R}-c(z-z_0)$ & 
                                                         \quad $g_{\rm R}(z_0)$ \\
\hline
10  & 1.3637 & 2.3199(2) & 3.0103(5)  & \quad 3.0105(1) & 3.0105(1) \\
28  & 1.575  & 2.3392(3) & 3.1132(6)  & \quad 3.0747(2) & \\
    & 1.578  & 2.3227(4) & 3.0818(8)  & \quad 3.0764(2) & \\
    & 1.5785 & 2.3200(3) & 3.0764(8)  & \quad 3.0765(2) & 3.0765(2)  \\
56  & 1.697  & 2.3201(4) & 3.0981(9)  & \quad 3.0979(2) & 3.0979(2)  \\
    & 1.70   & 2.3030(4) & 3.0650(9)  & \quad 3.0984(2) & \\
112 & 1.8074 & 2.3230(4) & 3.1155(9)  & \quad 3.1095(2) &  3.1095(2) \\
    & 1.81   & 2.3073(4) & 3.0843(9)  & \quad 3.1096(3) & \\
224 & 1.918  & 2.3159(5) & 3.1061(13) & \quad 3.1143(3) & \\
    & 1.9176 & 2.3178(5) & 3.1099(12) & \quad 3.1144(3) &  3.1145(3) \\
    & 1.9172 & 2.3212(5) & 3.1171(11) & \quad 3.1146(3) & \\
448 & 2.0276 & 2.3289(25)& 3.1355(60) & \quad 3.1178(15)& \\
    & 2.028  & 2.3204(10)& 3.1183(23) & \quad 3.1175(6) &  3.1175(6) \\
\hline
 10 & 0.8562 & 2.3226(8) & 3.1118(18) & \quad 3.1066(5) & \\
    & 0.8563 & 2.3217(8) & 3.1092(18) & \quad 3.1059(5) & \\
    & 0.8565 & 2.3202(4) & 3.1069(9)  & \quad 3.1064(2) &  3.1064(2) \\
    & 0.8567 & 2.3185(7) & 3.1034(17) & \quad 3.1063(5) & \\
    & 0.86   & 2.3050(10)& 3.0767(24) & \quad 3.1067(7) & \\
 20 & 1.00   & 2.3197(5) & 3.1135(9)  & \quad 3.1140(3) &  3.1140(3) \\
 40 & 1.1347 & 2.3235(7) & 3.1232(15) & \quad 3.1163(4) &  3.1163(4) \\
    & 1.14   & 2.2950(20)& 3.0666(44) & \quad 3.1166(11) & \\
    & 1.144  & 2.2762(22)& 3.0285(49) & \quad 3.1161(12) & \\
 80 & 1.25   & 2.4024(48)& 3.2827(112)& \quad 3.1179(30) & \\
    & 1.264  & 2.3298(16)& 3.1372(36) & \quad 3.1175(9) & \\
    & 1.2659 & 2.3199(9) & 3.1178(20) & \quad 3.1180(5) &  3.1180(5) \\
\hline
 28 & 1.56   & 2.3389(9) & 3.1041(22) & \quad 3.0663(7) & \\
    & 1.5636 & 2.3177(7) & 3.0639(17) & \quad 3.0685(4) &  3.0683(5) \\
    & 1.565  & 2.3099(8) & 3.0494(19) & \quad 3.0695(5) & \\
 56 & 1.664  & 2.3596(19)& 3.1702(46) & \quad 3.0900(14) & \\
    & 1.670  & 2.3190(9) & 3.0905(22) & \quad 3.0924(6) &  3.0923(7) \\
112 & 1.762  & 2.3242(10)& 3.1128(23) & \quad 3.1043(6) & \\
    & 1.7626 & 2.3211(6) & 3.1072(14) & \quad 3.1051(4) &  3.1051(5)\\
224 & 1.8465 & 2.3185(7) & 3.1077(16) & \quad 3.1107(4) &  3.1106(5) \\
\hline
\end{tabular}
    \caption{{}Measurements of $z=L/\xi(L)$, $g_{\rm R}$,
        $g_{\rm R}-c(z-z_0)$ (with $c=2.0$ and $z_0=2.32$), and 
        the extrapolated value $g_{\rm R}(z_0)$ for the O(3) nn action, 
        O(3) FP action and dodecahedron action, respectively. 
        Note that the error of $g_{\rm R}-c(z-z_0)$ is by a factor 
        $\sim 4$ smaller than that of $g_{\rm R}$. 
        The last column, $g_{\rm R}(z_0)$ is obtained by extrapolating
        the values of $g_{\rm R}-c(z-z_0)$ to $z=z_0$.}
    \label{grdata}
  \end{center}
\end{table}

\begin{figure}[htb]
\begin{center}
\leavevmode
\epsfxsize=100mm
\epsfbox{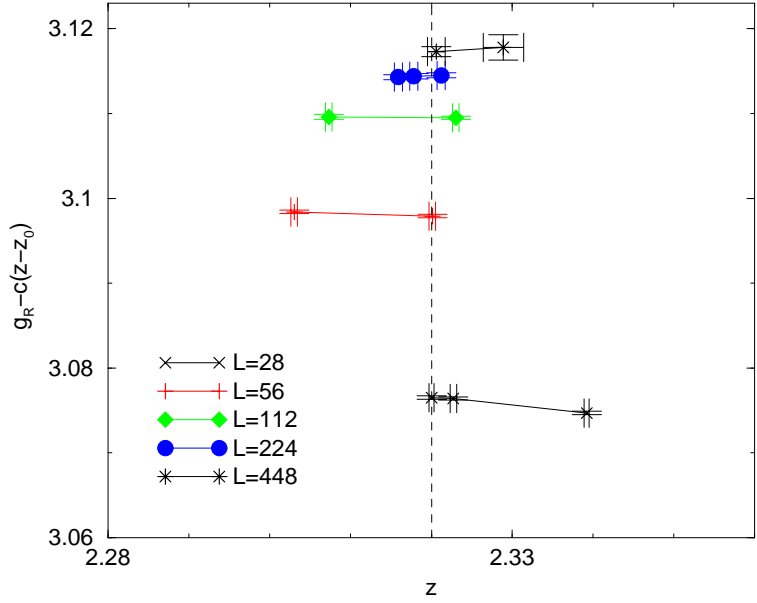}
\vskip -6mm
\end{center}
\caption{{}The quantity $g_{\rm R}-c(z-z_0)$ 
vs. $z$ for the standard O(3) action with $c=2.0$ and $z_0=2.32$.}
\label{fig:gro3z}
\end{figure}

Figure~\ref{fig:gro3z} shows the dependence of the quantity  
$g_{\rm R}-c(z-z_0)$ on $z$ around $z_0=2.32$ with $c=2$.
The value of $g_{\rm R}(z_0)$ is obtained by interpolating
these data to $z=z_0$.  
Since $z$ was quite close to the chosen value of $z_0$ this procedure 
introduces only a small, controllable error.
Similar plots for the other two actions are not
shown here but could be reconstructed from Table~\ref{grdata}.

\begin{figure}[htb]
\begin{center}
\vskip -3mm
\leavevmode
\epsfxsize=100mm
\epsfbox{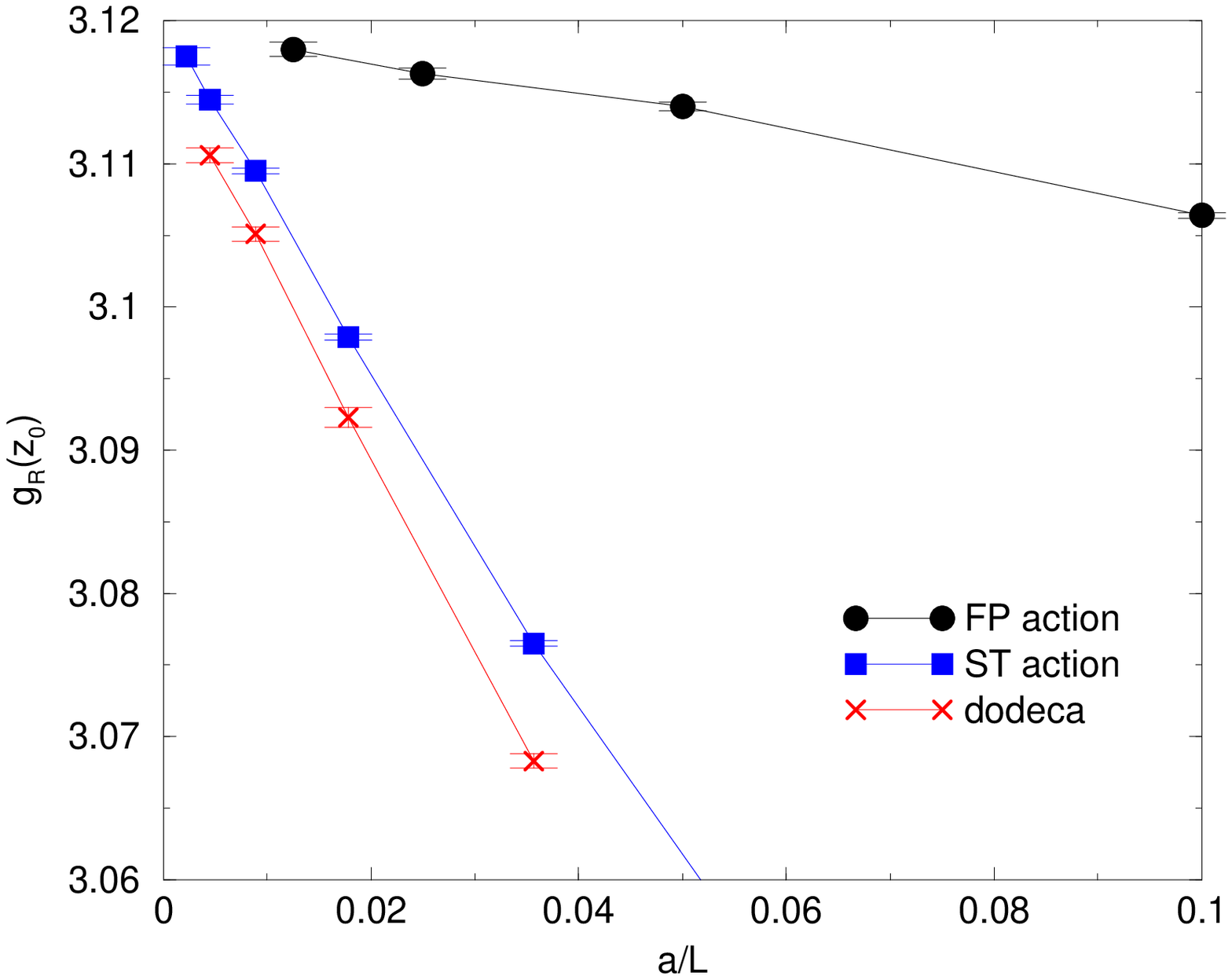}
\vskip -6mm
\end{center}
\caption{{}Cut-off dependence of $g_{\rm R}(z_0)$ (at $z_0=2.32$)
for the standard, FP and dodecahedron actions.
The data points are connected to guide the eye.}
\label{fig:gra}
\end{figure}

Figure~\ref{fig:gra} shows $g_{\rm R}(z_0)$ as the function of $a/L$ for
the three actions considered.

The statistical errors of $g_{\rm R}(z_0)$ at a given $a/L$ are quite 
small ($\sim 10^{-4}$ for the O(3) nn action).
However, the result of the extrapolation to the continuum limit 
of present data depends much more on the Ansatz used. 
Below we present fits with different functional forms,
together with the corresponding value of $\chi^2/{\rm dof}$:

\noindent Standard action:
\begin{alignat}{2}
& 3.1201(2)-1.224(9)a/L\,,                    & \qquad & (2.1) \nonumber \\
& 3.1205(4)-1.28(5)a/L + 1.1(1.1) (a/L)^2\,,  & \qquad & (2.2) \nonumber \\
& 3.1207(6)-1.1(1)a/L - 0.04(4)a/L \log(L/a)\,, & \qquad & (2.5) \nonumber \\
& 3.1168(3)+103(5)(a/L)^2-40.4(1.4)(a/L)^2\log(L/a)\,, & 
                                                \qquad & (4.1) \nonumber
\end{alignat}
\noindent FP action
\begin{alignat}{2}
& 3.1202(3)-0.137(4)a/L\,,                    & \qquad & (3.7) \nonumber \\
& 3.1186(7)-0.07(3)a/L - 0.5(2) (a/L)^2\,,    & \qquad & (2.5) \nonumber \\
& 3.1178(11)-0.23(4)a/L + 0.05(2)a/L \log(L/a)\,, & \qquad & (1.6) \nonumber\\
& 3.1180(5)+0.4(7)(a/L)^2-0.7(3)(a/L)^2\log(L/a)\,, & 
                                                \qquad & (1.6) \nonumber
\end{alignat}
\noindent dodecahedron action
\begin{alignat}{2}
& 3.1169(4)-1.36(2)a/L\,,                    & \qquad & (0.4) \nonumber \\
& 3.1168(9)-1.4(1)a/L - 0.5(3.1) (a/L)^2\,,  & \qquad & (0.9) \nonumber \\
& 3.1166(14)-1.4(3)a/L + 0.02(10)a/L \log(L/a)\,, & \qquad & (0.8)\nonumber\\
& 3.1129(6)+111(14)(a/L)^2-44(4)(a/L)^2\log(L/a)\,, & 
                                                \qquad & (0.4) \nonumber
\end{alignat}

Both from Figure~\ref{fig:gra} and these fits one concludes that an
$a/L$ term is needed in the Ansatz. 
(Note the large coefficients in the $a^2$ type fits in the standard
and dodecahedron action.)
We shall return to the question of cut-off effects in the last section.

The variation of the constant term in the fits represent the
systematic error of $g_{\rm R}(z_0)$. This is significantly larger than the
statistical error. 
Although it is somewhat arbitrary to represent this (theoretical) 
uncertainty by an error, based on these fits, we give the following 
estimates:
\begin{align}
g_{\rm R}^{\rm O(3),FP}(z_0)  &= 3.119(2) \,, \\
g_{\rm R}^{\rm O(3),nn}(z_0)  &= 3.120(2) \,, \\
g_{\rm R}^{\rm dodec,nn}(z_0) &= 3.117(2) \,.
\end{align} 

On the basis of standard universality arguments one expects
that the two O(3) actions give identical results in the 
continuum $a/L \rightarrow 0$. This is indeed the case.
The continuum limit of the dodecahedron model
is also in agreement within the errors given. It agrees
to 0.1\% with the O(3) results.

At this precision we are able to see, for the first time, cut-off
effects in the FP action results. A part of them might come from the 
unimproved operators as well.
At $L=28$
($\xi(L)\approx 10$) the cut-off effect is 0.004, a 0.1\% effect. 
At this correlation length the cut-off effect of the standard action 
is $\sim 10$ times larger.

We discuss here briefly how such high precision
could be reached with relatively small scale MC simulations.
This is due to several favorable circumstances.

\noindent
1) The quantities $\xi(L)$ and $g_{\rm R}$ are given directly by 
measuring sums over the entire lattice
without any need to extract the exponential fall-off of 
some correlation functions, hence without an extra systematic
error due to the fitting procedure.

\noindent
2) In the O(3) nn case we use ``fully improved'' estimators.
We build all clusters corresponding to all 3 components
(i.e. 3 cluster configurations) and apply the improved estimator
to all terms in 
$\langle ({\bf M}^2)^2 \rangle = 
\langle  M_1^4 + 2 M_1^2 M_2^2 + \ldots \rangle$.
This is allowed since for the standard O(3) action the 3
cluster distributions can be updated independently.
At $z\approx 2.3$ the use of improved estimator decreases 
the error by a factor $\sim 1.5$.
Unfortunately, there is no fully improved estimator for the FP action
(the action is not a sum of 3 terms each depending on one component only)
and for the dodecahedron (no $90^\circ$ rotation symmetry).
One still could apply the improved estimator to a single component
-- say $\langle  M_1^4 \rangle$ -- and restore the full quantity 
$\langle ({\bf M}^2)^2 \rangle$ using the symmetry.
However, this introduces an extra noise and for the present case 
(small $z$) it turns out to be worse than the standard estimator. 
Hence we used the improved estimator for these cases only
to check the self-consistency of the result
(in particular, to test the random number generator).

\noindent
3) As seen from eq.~(\ref{gR}), for large $z=L/\xi(L)$
there must be strong cancellation between the terms in the second
factor (the Binder cumulant).
Therefore the signal/noise ratio is much better for 
$z=z_0 = 2.32$ than, say, for $z\ge 7$.

\noindent
4) For small $z$ the quantities $g_{\rm R}$ and $z$ are strongly 
correlated with each other. 
The quantity $g_{\rm R} - c(z-z_0)$ with appropriate $c$
turns out to have much smaller fluctuations.
Choosing $c=2.0$ for $z_0=2.32$ the quantity $g_{\rm R}(z_0)$
defined in this way has an error $\sim 4$ times smaller than 
the error of $g_{\rm R}$ measured directly!

This way we could reach for $g_{\rm R}(z_0)$ in the nn O(3) model
a relative error smaller than $10^{-4}$, perhaps unprecedented in 
MC simulations of the non-linear $\sigma$-model.

\section{Gauge theories in $d=4$}

As opposed to the $d=2$ spin models discussed before, gauge models on
discrete non-Abelian subgroups of SU(2), or SU(3) have typically a
first order phase transition between the strong coupling and frozen
phases. The standard plaquette action, for example, with the largest
120-element subgroup $\tilde{Y}$ of SU(2) ($\tilde{Y}=Y \times Z_2$, 
where $Y$ is the symmetry group of dodecahedron) has a first order
phase transition at $\beta_W^{120} \approx 6$~\cite{Rebbi}. 
For the SU(2) plaquette
action approximate scaling starts already around $\beta_W^{\rm SU2}
\approx 2.7$ and at this coupling even the bare quantities of the SU(2)
and the $\tilde{Y}$ actions are very close to each other. One expects,
therefore that very large correlation lengths are reached with the
$\tilde{Y}$ action before the freezing transition occurs. For any
practical considerations this is the same situation as a second order
phase transition. 

In addition, the correlation length at the phase
transition point can be influenced by changing the form of the
action and we are not aware of any arguments which would exclude the
existence of a local action with a second order transition.
For the 1080-element subgroup of SU(3) with the plaquette
action the first order transition occurs at small correlation length
much before the scaling region\cite{Rebbi}. 
The transition is related to the large 
gap $\approx 1.4$ between the trace of the unit element and that of 
its nearest neighbor in group space. 
(Note that the corresponding gap is $\approx 0.4$ for the $\tilde{Y}$ 
subgroup of SU(2)).
If, by increasing $\beta$,
the plaquette expectation value is forced to take significantly larger
values than $3-1.4=1.6$, a first order transition occurs after which
most of the plaquette traces are at 3 and the system becomes frozen. 
This picture suggests to introduce additional loops into the action,
'distribute' the action value between them and so keeping the
plaquette value lower. Without systematic search we found several
actions which allow to go into the scaling region.

As an example, a recent parameterization of the SU(3) FP
action\cite{Nie}, when used over the 1080-element subgroup, remains
in the strong coupling phase down to an estimated $a \approx
0.03$fm. This resolution corresponds to $\beta_W^{\rm SU3}\approx 7$ for
the standard plaquette action. No serious numerical studies were
performed yet in SU(3) Yang-Mills theory at such small lattice
distances. 

\subsection{Tests with the largest non-Abelian subgroup of SU(2)}

We performed different tests on the 120-element $\tilde{Y}$ subgroup using
the plaquette action. No deviations from the SU(2) group predictions
were found. We have to remark, however that the precision
of these calculations is far below of that in the spin models presented
before.
 
\begin{figure}[htb]
\begin{center}
\leavevmode
\epsfxsize=100mm
\epsfbox{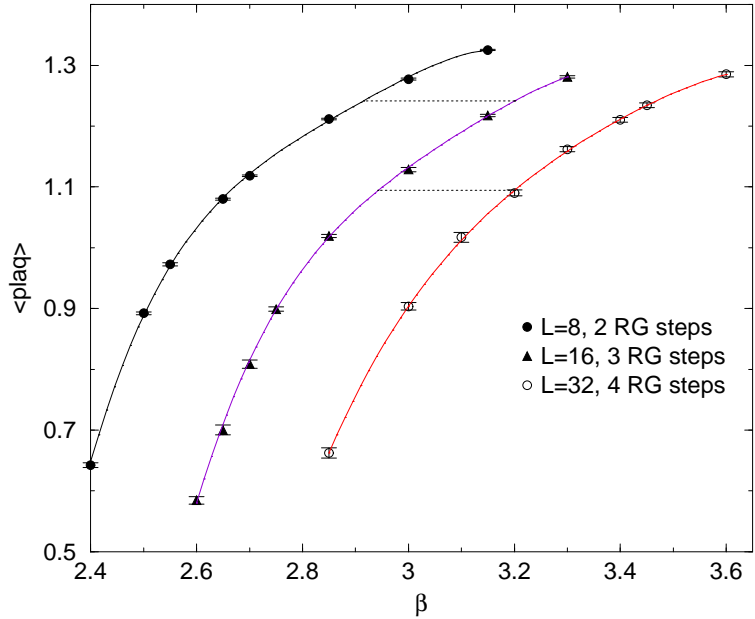}
\vskip -6mm
\end{center}
\caption{This figure offers a graphical way to determine the shift $\triangle
  \beta$ as the function of $\beta$. The change of scale between
  $\tilde{\beta}=\beta-\triangle \beta(\beta)$ and $\beta$ is a factor of 2:
  $a(\tilde{\beta}) = 2 a(\beta)$. The length of the horizontal line between
  the $L=16$ and $L=8$ curves gives the estimate $\triangle
  \beta(\beta=3.20)=0.285$ (3 vs. 2 blocking steps), that of between
$L=32$ and $L=16$ leads to $\triangle
  \beta(\beta=3.20)=0.257$ (4 vs. 3 blocking steps).} 
\label{fig:plaq}
\end{figure}

We made a standard two-lattice Monte Carlo renormalization group MCRG 
analysis in the coupling constant region $\beta \in
(2.6-3.5)$ \footnote{In this section the discrete group coupling
constant is denoted by $\beta$, while that of the SU(2) plaquette
action by $\beta^{\rm SU2}$.}
in order to test whether the underlying theory has a single
relevant parameter (as a Yang-Mills theory has) and to get quantitative
results on the scales, i.e. on $a(\beta)$, where $a$ is the lattice
unit. Configurations were generated on $32^4$ and $16^4$ lattices which were
blocked to $2^4$ lattices using 4 and 3 RG steps, respectively. A simple
Swendsen-type blocking \cite{Swendsen} was used and the block average was
projected back to the discrete subgroup after every block step. On the $2^4$
blocked configurations the plaquette and the length-6 twisted loop were
measured. They correspond to large smeared loops on the original
lattices. Generating the $32^4$ configurations at some $\beta$, one is
searching for an associated $\tilde{\beta} = \beta-\triangle
\beta $ where the $16^4$ configurations after 3 blocking steps give
the same loop expectation values as the $32^4$ configurations after 4
steps. The corresponding $\triangle \beta $ ( as the function of
$\beta $ ) gives the shift in the coupling which results in a factor
of 2 increase in the lattice unit $a$: $a(\tilde{\beta})= 2
a(\beta)$. In order to see the sensitivity on the number of blocking
steps, the procedure was repeated on $16^4$ versus $8^4$ lattices 
using 3 and 2 block steps, respectively.

\begin{figure}[htb]
\begin{center}
\leavevmode
\epsfxsize=100mm
\epsfbox{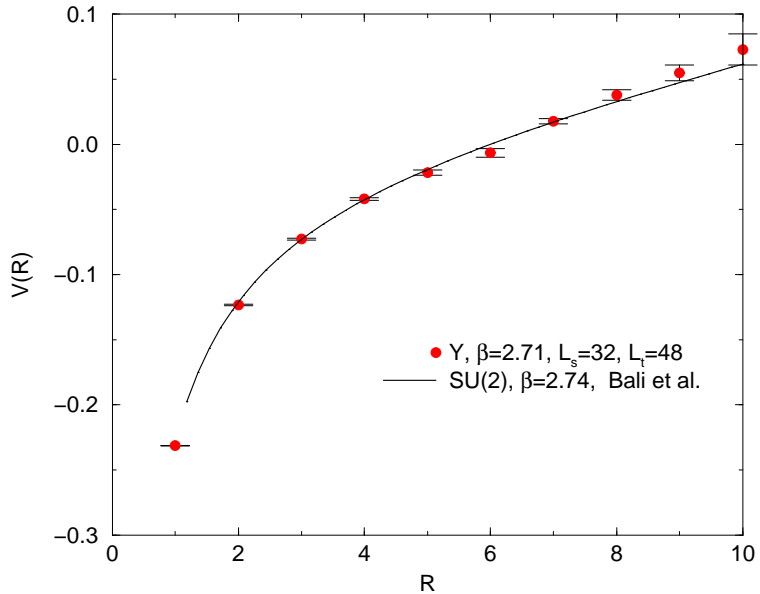}
\vskip -6mm
\end{center}
\caption{Matching the discrete 120-element $\tilde{Y}$ potential at 
$\beta_W^{120}=2.71$ with that of SU(2) at 
$\beta_W^{\rm SU2}=2.74$~\cite{Bali}. 
The scales are related as $a_{\rm SU2}(2.74)/a_{\tilde{Y}}(2.71)=0.84$.}
\label{fig:pot271}
\end{figure}

Fig~\ref{fig:plaq} gives the plaquette expectation values
and the figure caption explains how to get the $\triangle \beta$
shifts. The figure for the twisted loop expectation values looks similar.
The $\triangle \beta$ shifts from the plaquette and from the twisted loop are
consistent indicating that there is one relevant coupling only in this
problem. The $\triangle \beta$ values from this analysis 
(which are not far from the two-loop perturbation theory prediction for
SU(2)) suggest that very small lattice units can be obtained in this
formulation before approaching the freezing transition at $\beta \sim
6$. On the basis of this RG analysis we estimate that 
$a(\beta=2.72):a(\beta=2.95):a(\beta=3.20)=4:2:1$. We
are not able to give a reliable error estimate here due to the uncontrolled
extrapolation errors in the number of blocking steps.

\begin{figure}[htb]
\begin{center}
\leavevmode
\epsfxsize=100mm
\epsfbox{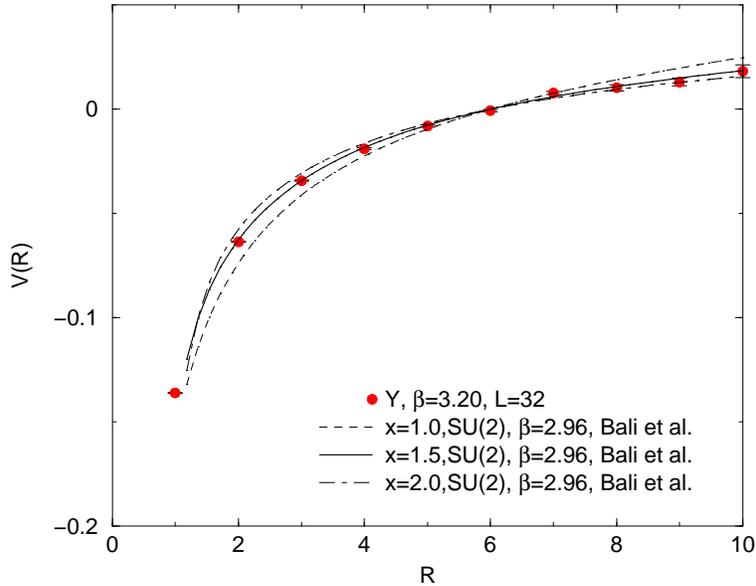}
\vskip -6mm
\end{center}
\caption{The same as Fig.~\ref{fig:pot271} at $\beta_W^{120}=3.20$ and 
$\beta_W^{\rm SU2}=2.96$~\cite{Bali}. The matching with 
$x=a_{\rm SU2}(2.96)/a_{\tilde{Y}}(3.20)=1.5$ is close to optimal, while
the curves with $x=1.0$ and 2.0 illustrate the (poor) sensitivity of the
matching on the ratio of scales.}
\label{fig:pot320}
\end{figure}

We measured the short distance part of the potential at
$\beta=2.71,2.96$ and 3.20 on $32^3 \times 48, 16^4$ and $32^4$
lattices, respectively, and compared the results with those obtained in SU(2)
at $\beta^{\rm SU2}=2.74$ and 2.96 by Bali et al.~\cite{Bali}. 
After matching the scales and the non-physical constant in the potential 
the subgroup results are consistent with the SU(2) curve. 
This is illustrated by Figs.~\ref{fig:pot271} and \ref{fig:pot320} for 
$\beta=2.71$ and 3.20, respectively. The quality of the matching is similar
at $\beta=2.96$. For the scales in the SU(2) and the subgroup model we
obtained $a_c(2.74)/a_d(2.71)=0.84, a_c(2.96)/a_d(2.96)=0.86$, and
$a_c(2.96)/a_d(3.20)=1.5$, where 'c' and 'd' refer to the continuous and
discrete groups, respectively. In Figure~\ref{fig:pot320} the 3 curves with
$x=a_c(2.96)/a_d(3.20)=1.0, 1.5, 2.0$ illustrate the sensitivity of the
matching on $x$. We did not attempt to give errors here since 
we did not check the finite volume effects and cut-off
effects in the short distance part of the potential. In addition, the
sensitivity of the matching on the scaling relations is rather weak. From
the scale relations above and $a_c(2.74)/a_c(2.96)=1.86$ \cite{Bali}
one obtains $a_d(2.71)/a_d(2.96)=1.92$, while the RG analysis gave 
$a_d(2.72)/a_d(2.95)=2$. Similarly, one obtains
$a_d(2.96)/a_d(3.20)=1.74$ and $a_d(2.95)/a_d(3.20)=2$ from potential
matching and RG analysis, respectively.

\section{Discussion on the cut-off effects}

We want to return to the question of cut-off effects in AF theories. 
Although there are no rigorous considerations beyond perturbation theory, 
it is a generally used standard assumption that close to the continuum 
limit $a^2$ powers with logarithms describe the cut-off effects in 
bosonic theories. Extrapolations of numerical data are done accordingly. 
The systematic Symanzik improvement program\cite{Sym} applied in a
non-perturbative environment is based on this assumption. 
We are perplexed by the results in Figure~\ref{fig:gra} and 
by the numbers in the $\chi^2$ fit. 
Actually, it has been observed earlier\cite{Bal} that at $z\gtrsim 5$
a linear fit for $g_{\rm R}$ in $a/L$ was preferred.
However, due to larger errors choosing this form was not compelling.
The leading order large $N$ result on $g_R$ in infinite volume has the
expected $\propto a^2$ (with logarithms) behavior\cite{largeN}. It
would be interesting to see what happens beyond leading order and in a
finite volume.
It would be important to check whether other observables, or
on-shell quantities like the spectrum show a similar behavior.
Obviously, the same questions in d=4 gauge theories are even more relevant.

\noindent {\bf Acknowledgement}

We benefited from numerous discussions with Adrian Patrascioiu and 
Erhard Seiler  on the similarity between the O(3) model and 
the corresponding discrete models.
We are indebted to Andrea Pellisetto and Alan Sokal for sending us the
para\-met\-rized form of the O(3) finite-size scaling function. We thank
Peter Weisz for valuable discussions, in particular for sharing
his insight on cut-off effects.
We thank also the organizers and participants of the exciting and
enjoyable Ringberg Workshop (April 2000), where we had the possibility to
present some the results of this paper. 
We are also indebted to
David Kaplan, Julius Kuti and Zolt\'an R\'acz for useful correspondence.

\eject

\end{document}